\documentclass[11pt]{article}
\usepackage{sao1}
\usepackage{graphicx}

\def\iue{\it IUE\/}
\def\oao{\it OAO\/}
\def\units{$10^{-10}\mathrm{erg}\hspace{0.8mm}\mathrm{s}^{-1}\mathrm{cm}^{-2}\mathrm{\AA}^{-1}$}
\begin{document}
%
%
%

\title{Spectrophotometric variability of the magnetic CP star
 $\alpha^{\rm 2}$~CVn
\thanks{Based on $INES$ data from the $IUE$ satellite.}}
\author{N. A. Sokolov}
\institute{Central Astronomical Observatory at Pulkovo,
St. Petersburg 196140, Russia \\ e-mail:~sokolov@gao.spb.ru}
\maketitle
\begin{abstract}
The spectrophotometric variability of the classical magnetic CP star
$\alpha^{\rm 2}$~CVn in the ultraviolet spectral region from 1150 to 3200~\AA\ is investigated.
This study is based on the archival {\it International Ultraviolet Explorer\/} data obtained at different phases of the rotational cycle.
The brightness of the star at spectral region from 3015 to 3138~\AA\ is constant over the period of rotation which means that the so called 'null wavelength region' exist on these wavelengths. Moreover, the minima values of the amplitude of light curves are reached in the spectral region at $\lambda\lambda$~1660 and 1900~\AA.
The monochromatic light curves in the 'pseudo-continuum' of this star
change their shape with wavelength.
All light curves with $\lambda$~$\leq$~2505~\AA\ have a similar shape, but
the monochromatic light curves with $\lambda$~$>$~2505~\AA\ shows the phase displacement of the minimum from 0.0 at $\lambda$~2505~\AA\ to 0.3 at $\lambda$~2993~\AA.


\end{abstract}

\section {Introduction}
The magnetic Chemically Peculiar (mCP) star $\alpha^{\rm 2}$~CVn (HD~112413,
HR~4915) was the first bright star ($m_{\rm v}$~=~2.90) classified as chemically peculiar of the SiHgEuCr~ type (Leckrone~1973) and is the prototypical magnetic CP star. This star shows periodic spectrum, magnetic field and light variations
with a period of 5.46939 days first established by Farnsworth~(1932).
In the visual spectral region the light curves for $\alpha^{\rm 2}$~CVn have been presented by Provin~(1953). In addition, twenty-one $UBV$ photometric observations of this star were also obtained by Pyper~(1969).

Molnar~(1973) investigated the ultraviolet photometric observations with
{\oao}-2 satellite for $\alpha^{\rm 2}$~CVn.
He showed that the light variations in the wavelength region shorter than
$\sim\lambda$~2960~\AA\ are generally in anti-phase to the light variations in
the visual region, although the shapes of light curves are different.
Moreover, the photometric light curves indicate that there are two important sources
of energy blocking in the far-UV region. Strong line blanketing by the rare earths
redistributes flux into the Paschen continuum, causing the major observed
photometric variations. In addition, a second source, which may be due to
a combination of continuous opacities and blanketing from the iron-peak and
rare-earth groups below $\lambda$~1600~\AA, apparently redistributes flux into
the region of the Balmer discontinuity.

Leckrone \& Snijders~(1979) have been studied the ultraviolet spectrophotometry obtained with $Copernicus$, {\oao}-2, $TD$-1 and a sounding rocket for
$\alpha^{\rm 2}$~CVn.
The authors compared the ultraviolet flux distribution at two phases 0.0 and 0.5
and they pointed out that: (1) at $\lambda$~$<$~1190~\AA, the flux varies with large
amplitude appears to result from enhanced continuous opacity sources, (2) variations
in the 1190--1365~\AA\ range are due primarily to lines of doubly ionized metals and
possibly of Si~{\sc ii}, and (3) from 1365 to 1800~\AA\ and from 2700 to 2900~\AA\
fluxes does not vary in any systematic way.
One can see that the variations of the ultraviolet fluxes are quite complex and it is necessary to investigate the variations of the ultraviolet fluxes at different phases of the rotational cycle for this star.

The light variability can be generally explained by the variable
abundance of several chemical elements observed in the atmospheres of
mCP~stars. Enhanced energy blocking decreases the flux in the far-UV region
where most of the lines of these elements are present.
The blocked flux re-emerges in the visual and the red parts of the spectrum.
Such an explanation is supported by the anti-phase relationship of light curves
in the far-UV and the visual spectral regions.
In addition, this relationship asserts the existence of the
'null wavelength region'  where the amplitude of light variations is
zero over the period of rotation
(see, for example, Leckrone~1974; Molnar~1975; Molnar et al.~1976; Jamar~1978).
Earlier photometric results obtained for some mCP~stars show
the existence of two or more 'null wavelength regions'
(Jamar~1977; Muciek et al.~1985).
Later, Sokolov~(2000, 2006, 2010) investigated the spectrophotometric variability
of two silicon mCP stars CU~Vir and 56~Ari in the far-UV spectral region, using
the archival {\iue} data. The author showed that in the case of CU~Vir the 'null wavelength region' is at $\lambda$~2000~\AA, but in the case of 56~Ari with more complex light curves 'null wavelength region' is not exists.
Detailed investigation of several additional stars is necessary to draw a definite conclusion about the mechanism of light variations in the spectra of mCP stars.
Another mCP star is $\alpha^{\rm 2}$~CVn for which there are enough
{\iue} data in order to investigate the spectrophotometric variability in the ultraviolet spectral region. The {\iue} Newly Extracted Spectra (INES) data from
the {\iue} satellite allows to investigate the variability of the monochromatic
light curves in the far-UV and near-UV spectral regions. The INES archive are
available from the INES Principal Center website {\rm http://ines.vilspa.esa.es}
or from the INES National Hosts Wamsteker~(2000).

In this paper, the low-dispersion spectra of mCP star $\alpha^{\rm 2}$~CVn are
analyzed in detail the variability of the monochromatic light curves in the far-UV and near-UV spectral regions, using the INES data from the {\iue} satellite.

\section {Observational Data}

\subsection{{\iue} spectra}

The following list of a series of two observations of $\alpha^{\rm 2}$~CVn
obtained with Short Wavelength Prime (SWP), Long Wavelength Redundant (LWR)
and Long Wavelength Prime (LWP) cameras was received from the INES
archive.
\begin{enumerate}
\item The first one contains 10 SWP spectra and 8 LWR spectra obtained
     in December 1981;
\item the second one contains 11 SWP and 7 LWP spectra obtained in March 1984.
\end{enumerate}
Additionally, four spectra (SWP~04812, SWP~04813, LWR~02674, LWR~04153) obtained
in October 1978 and March 1979 were received from the INES archive as well.
In all cases, the spectra were obtained through the large aperture (9.5$\arcsec$~$\times$~22$\arcsec$).

In the INES archive, each high-dispersion image has an associated
'rebinned' spectrum, which is obtained by rebinning the 'concatenated'
spectrum at the same wavelength step size (1.676~\AA/pixel for SWP camera and
2.669~\AA/pixel for LWR and LWP cameras) as low-resolution data
(Gonz{\'a}lez-Riestra et al.~2000).
This data set represents an important complement to the low-resolution
archive, and it is especially useful for time-variability studies.
The ultraviolet spectra used in this study are low-resolution echelle spectra
obtained with a resolution of about 6~\AA. Additionally, the 'rebinned' spectra
from high-dispersion images of $\alpha^{\rm 2}$~CVn were used, as well.

The inspection of the spectra of $\alpha^{\rm 2}$~CVn showed that the fluxes
of the SWP~27838 spectrum systematically lower and unsuitable for our purpose.
Additionally, the high-dispersion spectrum LWP~07741 was excluded, because
the exposure time is only 2.184 sec. Finally, we analyzed 22 SWP, 10 LWR and 6 LWP spectra, distributed quite smoothly over the period of rotation.

\subsection{The Period Variations of $\alpha^{\rm 2}$~CVn}

\begin{figure}[t]
\vspace{-0.2cm}
\centerline{\includegraphics[width=105mm]{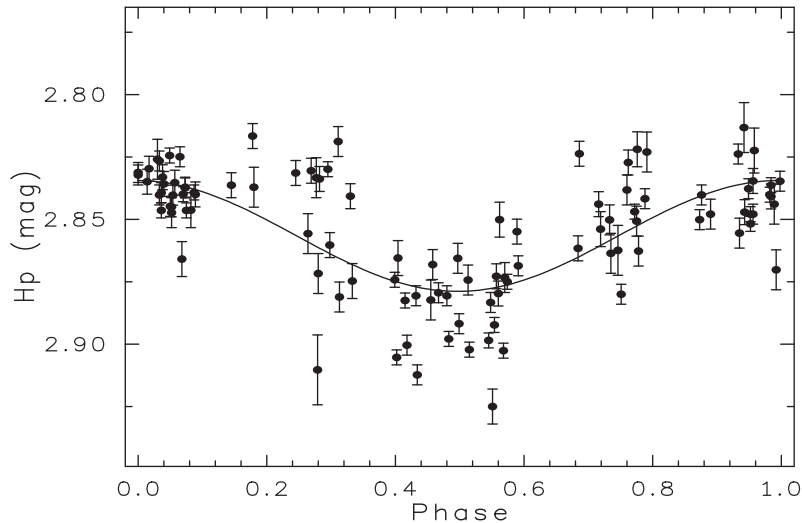}}
\caption{The Hipparcos photometry variations of $\alpha^{\rm 2}$~CVn.
 Solid lines denote the fit according to equation~\ref{equation_2}.}
\label{hip_ph}
\end{figure}

Since the beginning of the 20th century till now (e.g., Romanyuk et al.~2007)
the investigators use the ephemeris obtained of Farnsworth~(1932), which refers
to the phase of the maximum intensity of Eu~II spectral lines. On the other hand, some mCP stars displayed an increase their rotational periods
(Mikulasek et al.~2008).
Unfortunately, in literature there are not the photometric data, except of the $UBV$ photometric observations obtained by Pyper~(1969). Other publications concerning
the photometric observations of $\alpha^{\rm 2}$~CVn are not available as they are published only as plots. In order to check the trustiness of the ephemeris obtained
of Farnsworth~(1932) we used the photometric data from Hipparcos (ESA~1997).
The Hipparcos Epoch Photometry contains 107 measurements of $\alpha^{\rm 2}$~CVn
obtained between Julian dates: 2447896-2449018. The Hipparcos photometry (H$_{\rm p}$) are plotted in Fig.~\ref{hip_ph} versus the phase computed by means of Farnsworth~(1932) ephemeris:
\begin{equation}
{\mathrm JD}=2419869.720+{5^{\mathrm d}}\hspace{-0.9mm}.46939 E.
\label{ephemeris}
\end{equation}
We find a cosinusoidal variations of H$_{\rm p}$ with an amplitude of the order of 0.06~mag, although the scattering of H$_{\rm p}$ measurements around the fitted curve are significant. The mean standard deviations of the residual scatter around the fitted curve ($\sigma_{res}$) is equal to 0.018~mag.
Probably, such scattering of H$_{\rm p}$ measurements can be explained by the influence
of $\alpha^{\rm 1}$~CVn on the H$_{\rm p}$ data of $\alpha^{\rm 2}$~CVn at the passband of Hipparcos photometry.
The pair $\alpha^{\rm 2}$~CVn and $\alpha^{\rm 1}$~CVn is a visual binary with
a separation of 20$\arcsec$. The secondary is the spectral class F0~V and
$m_{\rm v}$~=~5.60.
In any case, the maximum and the minimum of the fitted H$_{\rm p}$ light curve
correspond to the phases 0.0 and 0.5, respectively. It is in the good agreement with
the $V$ light curve (see Fig.~1 of Pyper~1969).

\begin{figure*}[t]
\vspace{-0.2cm}
\centerline{\includegraphics[width=150mm, angle=0]{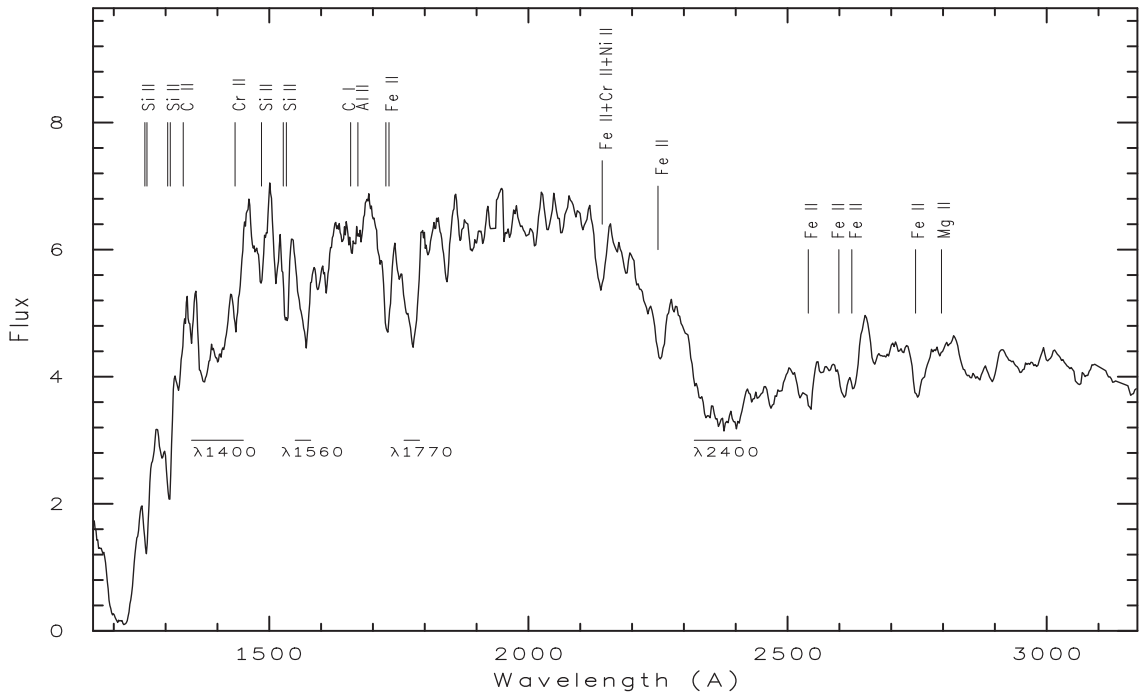}}
\caption{The average distribution of energy in {\units} for $\alpha^{\rm 2}$~CVn.
 The prominent spectral lines and features are shown by vertical and
 horizontal lines, respectively.}
\label{mean}
\end{figure*}

\section {Data Analysis}

To analyze the {\iue} spectra of $\alpha^{\rm 2}$~CVn we used the linearized least-squares method. An attempt was made to describe the light curves in a
quantitative way by adjusting a Fourier series.
The method has already applied to the {\iue} data of two silicon mCP stars
CU~Vir and 56~Ari and has shown the very good descriptions of the monochromatic
light curves (Sokolov~2000, 2006, 2010).
In the case of $\alpha^{\rm 2}$~CVn the observations were fitted by a simple
cosine wave:
\begin{equation}
F(\lambda,\,t)=A_{0}(\lambda) +
A_{1}(\lambda)\cos(2\pi(t-T_{0})/P +\phi(\lambda)),
\label{equation_2}
\end{equation}
where $F(\lambda,\,t)$ is a flux for the given $\lambda$ and the $t$ is Julian date of the observation.
The $T_{0}$ and $P$ are zero epoch and rotational period of the ephemeris, respectively.
The coefficients $A_{0}(\lambda)$ of the fitted curves
define the average distribution of energy over the cycle of the variability while
the coefficients $A_{1}(\lambda)$ define the semi-amplitude of the flux variations
for the given $\lambda$. From several scans distributed over the period one can
produce light curves at different wavelengths. This procedure can be partially accounted for by considering that within the accuracy of the measurements a cosine wave appear to be generally adequate to describe the monochromatic light curves in
the ultraviolet spectral region.

The least-squares fit was applied to the separate {\iue} monochromatic light curves. An error analysis showed that the uncertainties in the coefficients $A_{0}(\lambda)$ and $A_{1}(\lambda)$ of the fitted curves not more 0.05 and 0.07, respectively. 
Although, the standard deviations of the residual scatter around the fitted curves ($\sigma_{res}$) varies from 0.03 to 0.18 in the investigated wavelengths. 
The maximal errors of the coefficients $A_{0}(\lambda)$ and $A_{1}(\lambda)$ as well as in $\sigma_{res}$ are in the blue and red parts of {\iue} spectra.
Probably, it is connected with uncertainties of the fluxes in both ends of spectra, as presented in INES database.
Thus, we limited our investigation to the wavelength region from 1150 to 1950~\AA\
and from 1950 to 3200~\AA\ for far-UV and near-UV spectral regions, respectively.
In order to minimize the uncertainties in the coefficients of the fitted curves,
the light curves were determined by averaging three nearest fluxes for a given $\lambda$:
\begin{equation}
 F(\lambda) = \frac{F(\lambda-\lambda_{step}) + F(\lambda) + F(\lambda+\lambda_{step})}
 {3},
\label{equation_3}
\end{equation}
where $\lambda_{step}$ is equal 1.676~\AA\ for SWP camera and is equal 2.669~\AA\
for LWR and LWP cameras. As far as the errors in $F(\lambda)$ are concerned, we computed them by taking into account the the errors in the fluxes as presented in INES $Catalog$, according to the standard propagation theory of errors.

\begin{figure*}[t]
\vspace{-0.2cm}
\centerline{\includegraphics[width=165mm, angle=0]{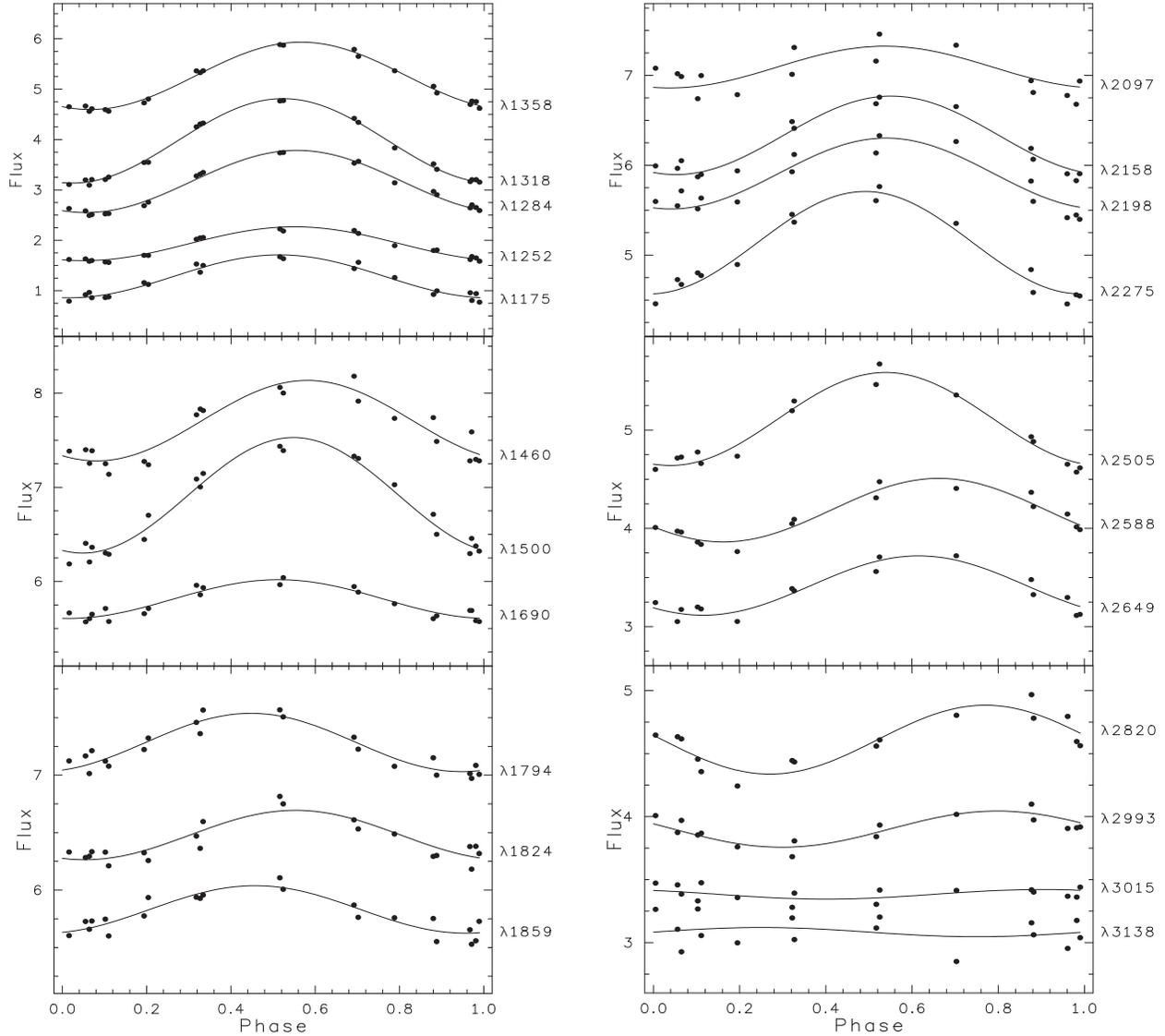}}
\caption{Phase diagrams of the monochromatic light curves in {\units} for
 $\alpha^{\rm 2}$~CVn. To avoid overlapping the vertical shift on the constant
 value of some curves was used (see text).
 Solid lines denote the fit according to equation~\ref{equation_2}.}
\label{mono_light}
\end{figure*}

\section {Monochromatic Brightness Variations in the Pseudo--Continuum}

Figure~\ref{mean} displays the average energy distribution $(A_{0}(\lambda))$
of  $\alpha^{\rm 2}$~CVn over the cycle of the variability in the spectral region from 1150 to 3200~\AA.
First of all, it is necessary to fix the continuum in the low-dispersion
{\iue} data. This is very difficult in the UV due to the lines
crowding. Nevertheless, one can find some high flux points located
at the same wavelength in several spectra of $\alpha^{\rm 2}$~CVn.
It should be noted that such choice of the continuum might be a
'pseudo-continuum'. However, there is no chance to reach the true continuum,
if it occurs at high flux points.
From several scans distributed over the period of rotation one can produce
light curves in different wavelengths. The light curves discussed below
will be called 'monochromatic', although they were determined by
averaging three nearest fluxes for a given $\lambda$ according to equation~\ref{equation_3}.
Several monochromatic light curves in the 'pseudo-continuum' at different
wavelengths were formed. The examples of light curves together with the fitted
cosine curves are shown in Fig.~\ref{mono_light}.
Note the vertical scales differ for each part of the figure. In order to exclude
overlapping of some curves the vertical shift on the constant value was used.
In this way, the curves at $\lambda\lambda$~2649, 2993 and 3015~\AA\ were
shifted down to the values of -1.5, -0.5 and -1.0~$\times$~{\units},
respectively. On the other hand, the curves at $\lambda\lambda$~2097 and 2505~\AA\
were shifted up to the values of +0.5 and +1.0~$\times$~{\units}, respectively.

The monochromatic light curves in the 'pseudo-continuum' of $\alpha^{\rm 2}$~CVn
change their shape with wavelength.
All light curves with $\lambda$~$\leq$~2505~\AA\ have a similar shape: a minimum of
the flux at phase 0.0 and a maximum of the flux at phase 0.5.
On the other hand, the monochromatic light curves with $\lambda$~$>$~2505~\AA\
shows the phase displacement of the minimum from 0.0 at $\lambda$~2505~\AA\ to 0.3 at $\lambda$~2993~\AA. The maximum also moves from the phase 0.5 at $\lambda$~2505~\AA\ to the phase 0.8 at $\lambda$~2993~\AA. Moreover, the amplitude of light variations decreases with increasing wavelength and, as a result, at $\lambda$~$>$~2993~\AA\ there is the 'null wavelength region', where the amplitude of light variations is zero over the period of rotation. Our result is in the agreement with previous investigation of Molnar~(1973) obtained from the ultraviolet photometric observations with {\oao}-2 satellite for $\alpha^{\rm 2}$~CVn.
It is necessary to note that the amplitude of light variations can also changes in the spectral region with $\lambda$$\leq$~2505~\AA, as illustrated by Fig.~\ref{mono_light}.

\begin{figure*}[t]
\vspace{-0.2cm}
\begin{center}
\centerline{\includegraphics[width=150mm, angle=0]{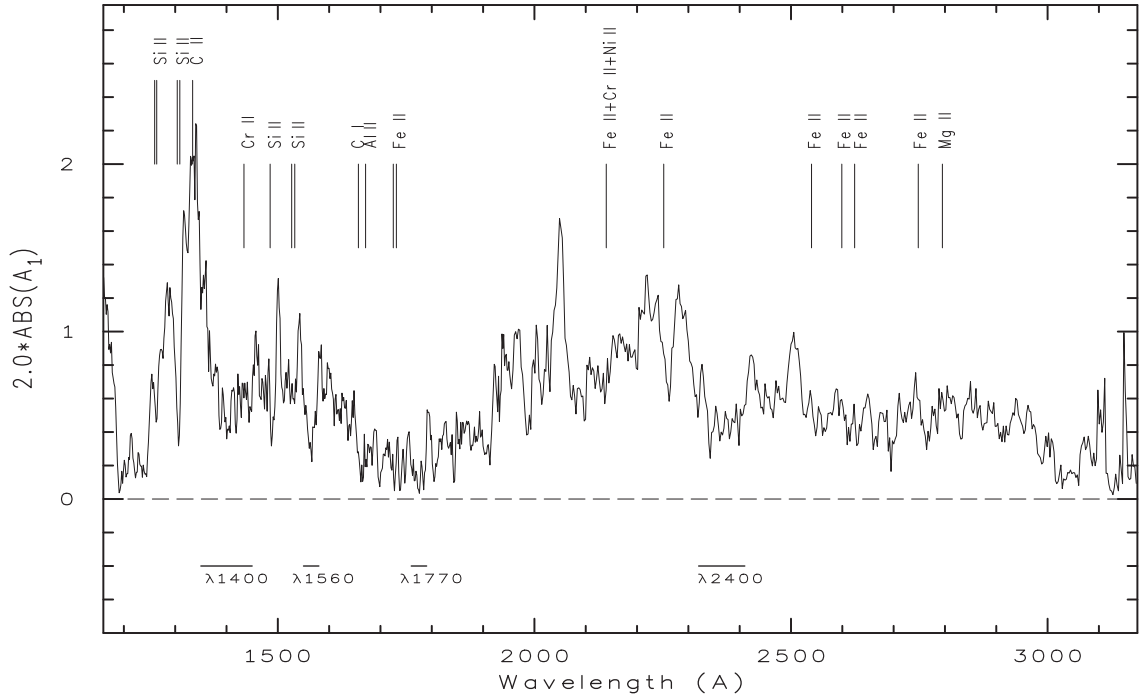}}
\caption{The amplitudes of the monochromatic light curves in {\units} for
 $ \alpha ^ {\rm 2} $ ~CVn. The prominent spectral lines and features are shown
 by vertical and horizontal lines, respectively.}
\end{center}
\label{amplitude}
\end{figure*}

\section{The Null Wavelength Regions}

The situation with position of 'null wavelength region' in the spectrum of $\alpha^{\rm 2}$~CVn is enough mysterious. Thus, Molnar~(1973) had determined
the position of the 'null wavelength region' at $\sim\lambda$~2960~\AA.
On the other hand, Leckrone \& Snijders~(1979) find that 'null wavelength region'
extends short wavelength to about 2700~\AA. Moreover, they have noted that from 1365 to at least 1800 \AA\  and from 2700 to at least 2900 \AA, $\alpha^{\rm 2}$ ~CVn  does not vary in any systematic way.

The INES data from the {\iue} satellite allows to estimate more accurately the position of the 'null wavelength region' in the spectrum of this star.
In order to investigate behavior of the monochromatic light curves in the 'pseudo-continuum' of $\alpha^{\rm 2}$~CVn we used
the coefficients $A_{1}(\lambda)$ (semi-amplitude of the flux variations).
Figure~\ref{mono_light} presents the dependence of the amplitude of the light curves from wavelength. One can see from Fig.~\ref{mono_light} that the variations of the amplitudes of
light curves is quite complex. The maximal values of the amplitudes are reached in
the spectral regions $\lambda\lambda$~1281--1294, 1314--1360, 1500, 2048, 2219,
2275, 2420 and 2505~\AA. On the other hand, the minimal values of the amplitudes
are in the spectral region from 1660 to 1900~\AA\ as well as in the region with $\lambda$~$>$~2993~\AA. Also, the minimal values of the amplitude are reached
at the cores of the large features at $\lambda\lambda$~1560 and 1770 \AA\ and at the cores of the strong Si~II resonance lines at $\lambda\lambda$~1260--64, 1304--09 and 1485~\AA. Moreover, the minimal values of the amplitude are at the cores of the Fe~II depression at $\lambda\lambda$~1725--31 and 2250~\AA.

\begin{figure}[t]
\vspace{-0.2cm}
\centerline{\includegraphics[width=90mm]{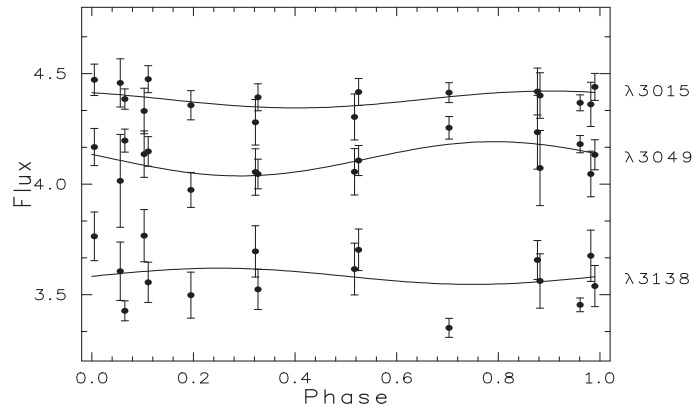}}
\caption{Phase diagrams of the monochromatic light curves in {\units} in
 the near-UV spectral region for $\alpha^{\rm 2}$~CVn.
 Solid lines denote the fit according to equation~\ref{equation_2}.}
\label{null_reg}
\end{figure}

In order to establish the position of the 'null wavelength region' in
the spectral region with $\lambda$~$>$~2993~\AA\ the amplitudes of the light variations in the 'pseudo-continuum' have been used. The inspection of the values
of amplitudes at this spectral regions showed that within the errors of measurements the fluxes are not varies in the spectral region from 3015 to 3138 \AA,
as illustrated by Fig.~\ref{null_reg}. In other words, the brightness of the star at this spectral region is constant over the period of rotation which means that the so called 'null wavelength region' exist on these wavelengths.

\section{Discussion}

For the first time, Molnar~(1973) investigated the photometric variability
of $\alpha^{\rm 2}$~CVn in the far-UV spectral region using the photometric data
obtained with {\oao}-2 satellite. Unfortunately, it is very difficult to compare
the phase diagrams obtained from the photometric data with our phase diagrams,
because the {\oao}-2 photometric  filters had FWHM~$\ga$~200~\AA\
(Code et al.~1970).
On the other hand, the {\iue} satellite performed UV spectrophotometry in the low-resolution mode with spectral resolution $\sim$6~\AA.
Although, there is some evidence for the phase displacement of minimum of the flux
at $\lambda\lambda$~1332, 1430 and 1554~\AA\ by obtained from the photometric data
in the far-UV spectral region.
These have a very strong minimum of the flux at phase 0.1 and a secondary minimum at phase 0.5. However, this latter feature is not very evident at $\lambda$~1554~\AA.
Perhaps the minimum of the flux at phase 0.1 is related to the maximum of the light curves at phase 0.1 for the light curve at $\lambda$~3317~\AA\ and the light curves
in the $U$ and $B$ filters (see Fig.~1 and 2 of Molnar~1973).
Molnar~(1973) had indicated that there are two important sources in the blocking
of the emergent flux in the far-UV spectral region. Strong blanketing by the rare-earths elements (REE) redistributes flux into the visual spectral region. In addition, a second source, which may be due to a combination of continuous opacities and line blanketing from the iron-peak and rare-earth groups below $\lambda$~1600~\AA. Leckrone \& Snijders~(1979) noted that the complex shapes of Molnar's far-UV light curves can be understood in terms of the competition between Cr, Fe and Si lines, which vary out of phase with the REE lines, and Ti, Mn and V, which vary approximately in phase with the REE lines.

Probably, an uneven surface distribution of chromium, silicon and iron mainly influence on the flux redistribution from the far-UV to the visual spectral regions in the spectrum of $\alpha^{\rm 2}$~CVn, although an additional sources of opacity can be involved.
It is not new that chromium, silicon and iron influence on the flux redistribution in the spectra of mCP stars. Khan \& Shulyak~(2007) have studied the effects of
individual abundance patterns on the model atmospheres of mCP stars. They have shown
that the group of elements which produce large changes in the model atmosphere
structure and energy distribution mainly consists of silicon, iron and chromium.
It should be noted that lines of these elements are widely presented in spectra
of mCP stars in the far-UV spectral region.
Recently, Krticka et al.~(2009) have simulated the light variability of the star
HR~7224 using the observed surface distribution of silicon and iron.
They have concluded that a promising explanation for the light variations in mCP stars is a flux redistribution through line and bound-free transitions combined
with the inhomogeneous surface distribution of various elements.

Our investigation indicate that the variations of the monochromatic light curves in the 'pseudo-continuum' of  $\alpha^{\rm 2}$~CVn are more complex.
For example, the monochromatic light curves with $\lambda$~$>$~2505~\AA\ shows
the phase displacement of the minimum.
Moreover, the minimal values of the amplitude are reached
at the cores of the large features at $\lambda\lambda$~1560 and 1770 \AA\ and at the cores of the strong Si~II resonance lines at $\lambda\lambda$~1260--64, 1304--09 and 1485~\AA. Also, the minimal values of the amplitude are at the cores of the Fe~II depressions at $\lambda\lambda$~1725--31 and 2250~\AA.
Possibly, an additional investigation the variations of the flux at the cores of the large features and spectral lines will help to understand such behavior of the light curves in the spectrum of $\alpha^{\rm 2}$~CVn.
It are given in the next paper by Sokolov~(2011).

\section {Conclusions}

The archival {\iue} spectrophotometric observations of $\alpha^{\rm 2}$~CVn
have permitted to analyze the light variations in the spectral region from
1150~\AA\ to 3200~\AA\ at various wavelengths.
The monochromatic light curves in the 'pseudo-continuum' of $\alpha^{\rm 2}$~CVn
change their shapes with wavelength.
All light curves with $\lambda$~$\leq$~2505~\AA\ have a similar shape: a minimum of
the flux at phase 0.0 and a maximum of the flux at phase 0.5.
On the other hand, the monochromatic light curves with $\lambda$~$>$~2505~\AA\
shows the phase displacement of the minimum from 0.0 at $\lambda$~2505~\AA\
to 0.3 at $\lambda$~2993~\AA.
The brightness of the star at spectral region from 3015 to 3138~\AA\ is constant
over the period of rotation which means that the so called 'null wavelength region' exist on these wavelengths.

\end{document}